\documentclass[showpacs,preprintnumbers,10pt,twocolumn]{revtex4}
\usepackage{dcolumn}
\usepackage{bm}
\usepackage{graphicx}

\begin{document}

\title{Resonance-inclined optical nuclear spin polarization of liquids in diamond structures}
\author{Q. Chen$^{1,2,\dag,*}$, I. Schwarz$^{1,2,\dag}$, F. Jelezko$^{2,3}$, A. Retzker$^{4}$, and M.B. Plenio$^{1,2}$ }
\affiliation{$^{1}$ Institut f\"{u}r Theoretische Physik, Albert-Einstein-Allee 11, Universit\"{a}t Ulm, 89069 Ulm, Germany\\
 $^{2}$ IQST,  Albert-Einstein-Allee 11, Universit\"{a}t Ulm, 89069 Ulm, Germany\\
 $^{3}$ Institut f\"{u}r Quantenoptik, Universit{\"a}t Ulm, 89073 Ulm, Germany\\
$^{4}$ Racah Institute of Physics, The Hebrew University of Jerusalem, Jerusalem, 91904, Israel\\
$^{\dag}$ These authors contributed equally to this work}
\email{qiong.chen@uni-ulm.de}

\begin{abstract}
Dynamic nuclear polarization (DNP) of molecules in a solution at room temperature has potential
to revolutionize nuclear magnetic resonance spectroscopy and imaging. The prevalent methods for
achieving DNP in solutions are typically most effective in the regime of small interaction
correlation times between the electron and nuclear spins, limiting the size of accessible
molecules. To solve this limitation, we design a mechanism for DNP in the liquid phase that is
applicable for large interaction correlation times. Importantly, while this mechanism makes use
of a resonance condition similar to solid-state DNP, the polarization transfer is robust to a
relatively large detuning from the resonance due to molecular motion. We combine this
scheme with optically polarized nitrogen vacancy (NV) center spins in nanodiamonds to design a
setup that employs optical pumping and is therefore not limited by room temperature electron
thermal polarisation. We illustrate numerically the effectiveness of the model in a flow cell
containing nanodiamonds immobilized in a hydrogel, polarising flowing water molecules 4700-fold
above thermal polarisation in a magnetic field of 0.35 T, in volumes detectable by current NMR
scanners.
\end{abstract}
\maketitle

\textit{Introduction}: Nuclear spin hyperpolarization, i.e. a population difference between
the nuclear spin states that exceeds significantly the thermal equilibrium value, is a key emerging
method for increasing the sensitivity of nuclear magnetic resonance (NMR)~\cite{Bloch,Bloch2}
which is proportional to the sample polarisation. By enhancing the magnetic resonance signals
by several orders of magnitude, a wide range of novel applications in biomedical sciences are made possible,
such as metabolic MR imaging~\cite{law2002high} or characterization of molecular chemical composition~\cite{luchinat2008nuclear,bertini2005nmr}. Dynamic nuclear polarization (DNP), by which electron spin polarization is transferred to nuclear spins, is one of the promising
methods to reach such a large enhancement of the signal.
 Nuclear spin hyperpolarization, i.e. a population difference between
the nuclear spin states that exceeds significantly the thermal equilibrium value, is a key emerging
method for increasing the sensitivity of nuclear magnetic resonance (NMR)~\cite{Bloch,Bloch2}
which is proportional to the sample polarisation. By enhancing the magnetic resonance signals
by several orders of magnitude, a wide range of novel applications in biomedical sciences are made possible,
such as metabolic MR imaging~\cite{law2002high} or characterization of molecular chemical composition~\cite{luchinat2008nuclear,bertini2005nmr}. Dynamic nuclear polarization (DNP), by which strong electron spin polarization is transferred to nuclear spins, is one of the promising
methods to reach such a large enhancement of the signal.
Over the past decades, one of the outstanding challenges is the hyperpolarization of molecules in a solution~\cite{nesmelov2004aqueous,abragam,mccarney2008spin1,mccarney2008spin2,ebert2012mobile,anderson1962influence,Gabl}.
In typical solutions, resonance-based polarization mechanisms commonly used in solid-state systems~\cite{abragam1978,wenckebach2008solid} are not effective due to the averaging of the electron-nuclear anisotropic interaction of the molecules by their rapid motion. Thus, cross-relaxation
mechanisms such as the Overhauser effect~\cite{abragam,anderson1962influence,Gabl} are
typically applied to realise polarization of fluids under ambient conditions. So far,
the largest hyperpolarization is achieved for fast diffusing small molecules
and is severely limited by the low thermal electron polarization in moderate magnetic fields.
For example, for TEMPO, trityl or biradical systems, the electron spin polarisation amounts
to only 0.08\% percent for a magnetic field of $B=0.36$ T~\cite{mccarney2008spin1,mccarney2008spin2}.

\begin{figure}
\center
\includegraphics[width=3.2in]{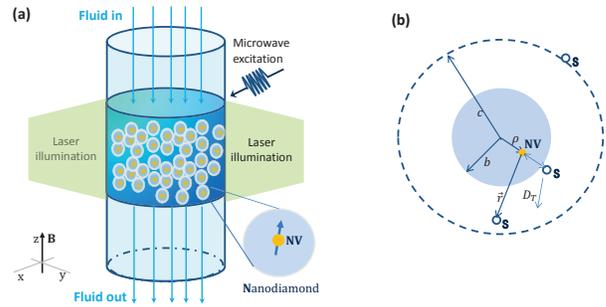}
\caption{(a) Schematics of the setup for polarisation of nuclear spin in solution via NV centers in nanodiamonds. Water is pumped into a channel containing a hydrogel cell with the immobilized nanodiamonds for continuous DNP polarisation. Laser illumination is used for polarising the NV spins and WM irradiation facilitates the transfer to nuclear spins in surrounding molecules. The hyperpolarized water then flows downstream for NMR detection.
A permanent magnetic field $B$ is applied along the $z$-axis, which defines the coordinate
system. (b) Illustration of the model parameters.}
\label{setup}
\end{figure}

In this work, we demonstrate that the use of optically hyperpolarized
electron spins offers an exciting possibility for overcoming the limitation
on the degree and rate of electron polarization. Specifically, nitrogen vacancy
(NV) centers in nanodiamonds make an excellent candidate as optically pumped
hyperpolarisation agents. The unique optical properties of the negatively
charged nitrogen-vacancy center in nanodiamonds allow for over 90\% electron
spin polarization to be achieved in less than a microsecond by optical
pumping \cite{Manson2006} while exhibiting a relaxation time in the millisecond
range even at room temperature~\cite{jelezko}. Methods for the
creation of high polarization in $^{13}C$ nuclear spins inside of bulk diamond
have already been developed theoretically and demonstrated experimentally~\cite{chen,Cai,Cai2,CaiRJP13,London,King,Bretschneider}.

However, for NV centers the interaction correlation time $\tau_c$
with nuclear spins in surrounding molecules is atypically large, as $\tau_c$ scales
with the square of the minimum distance with the nuclear spins, which is very large
(a few nanometers) for NV centers in nanodiamonds. Thus, NV centers cannot typically be used
with standard Overhauser cross-relaxation protocols. Here, we present a new theoretical
framework for polarizing molecules for systems with a large correlation time.
Specifically, we apply resonance-based schemes, such as the solid effect, where under
continuous microwave (MW) radiation the electron spin is driven to approach resonance
with the nuclear spin of choice. Importantly, we will proceed to demonstrate that
this scheme remains robust even in the presence of molecular motion and, unlike
solid-state polarization methods, is tolerant to a relative large detuning from
the resonance frequency.
\begin{figure*}
\center
\includegraphics[width=2.4in]{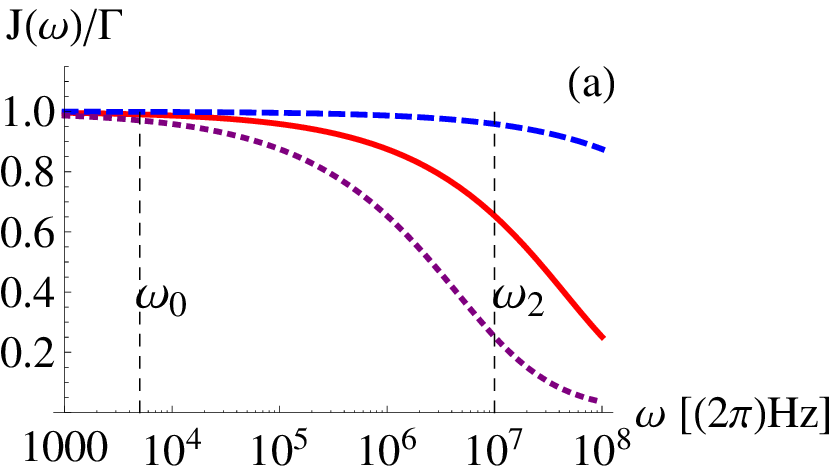}
\includegraphics[width=2.6in]{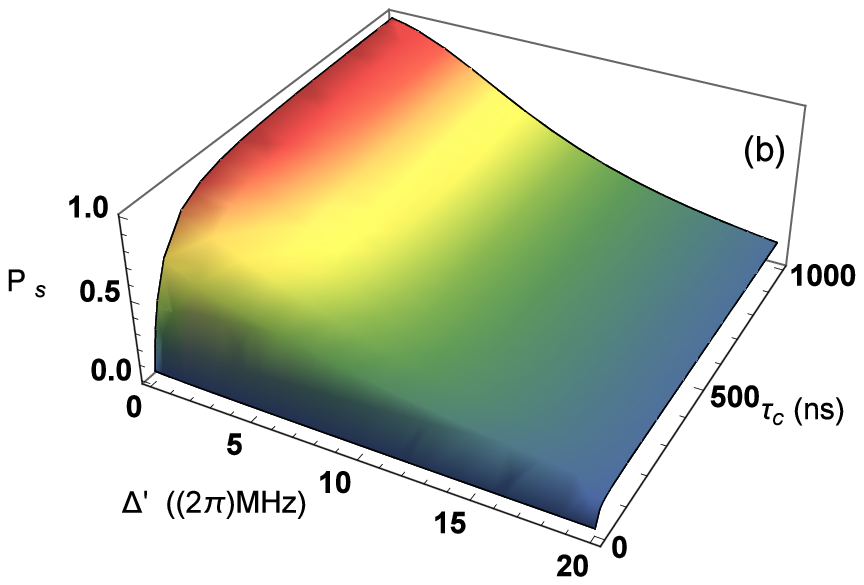}
\caption{(a) Spectra density function $J(\omega)$ as a function of frequency $\omega$ and interaction correlation time $\tau_c$ . The blue dashed, red solid, and purple dotted lines correspond to $\tau_c=0.1$ ns, 10 ns and 100 ns respectively. (b) The steady-state polarization of the nuclear spins. $\omega_S=(2\pi)16$ MHz, $\epsilon_0=\Omega=(2\pi)8\sqrt{2}$ MHz, and $\Delta'=\epsilon(\theta)-\epsilon_0$ is the detuning from the resonance point. 
}
\label{J0}
\end{figure*}

We then proceed to consider a specific setup that allows to realise this protocol, where
the nanodiamonds are immobilized in a hydrogel inside a flow channel, increasing
$\tau_c$, as well as limiting the nuclear spin relaxation due to electron spins to the
polarization region. Due to the optical NV polarization, and the efficiency of our
scheme in the large $\tau_c$ regime, a very high nuclear spin polarization is achieved
for volumes that are detectable in current NMR scanners.


\textit{Spin polarisation via resonance-inclined transfer (SPRINT) --} We consider
the polarization transfer induced by magnetic dipole-dipole coupling between an
electron spin and all the surrounding solvent spins. Suppose as shown in Fig.
\ref{setup}(a) that the electron spin is located in the center of a nanoparticle,
assuming the radius of the small solvent molecule to be negligible, the distance
of closest approach of the electron and nuclear spins, $b$, coincides with the radius
of the nanoparticle. All the solvent molecules are diffusing in a solution with the
translational diffusion coefficient $D_T$. The interaction between electron and solvent
spins evolves in time with a characteristic correlation time $\tau_c=b^2/D_T$. $c$
and $\rho$ in Fig. \ref{setup}(b) are related to absorbing boundary and off-center
effect, respectively, which are negligible in our case, see supplementary information (SI)~\cite{SI}.

In resonance-based DNP schemes, continuous MW radiation is applied to the electron
spin~\cite{abragam1978,wenckebach2008solid}, thus creating an ``effective frequency"
of the electron spin $\omega_{E}$ in the rotating frame. When the energy difference
between the effective frequency and the Larmor frequency $\omega_S$ of a specific
nuclear spin species vanishes ($\omega_{0} = \omega_{E}-\omega_S\rightarrow0$) a
Hartmann-Hahn (H-H) resonance is achieved~\cite{Hartman}) and energy-conserving
flip-flop transitions can occur between the electron and nuclear spins. Energy non-conserving flip-flip transitions are suppressed due to large energy mismatch
compared with the effective coupling between the electron and nuclear spin
($\omega_{2} = \omega_{E}+\omega_S\gg g$). This difference between the flip-flop
and flip-flip transitions leads to a net polarization transfer from electron
spins to nuclear spins~\cite{SI,Solomon}.

However, in the liquid phase, due to fast molecular diffusion, the nuclear spins
interact with the electron spin within a finite correlation time $\tau_c$. Perturbatively, 
we calculate the transition probabilities $W_i$ ($i=0,2$) of the flip-flop transition
rate $W_0=\alpha_0J( \omega_0)$ and flip-flip transition rate
$W_2=\alpha_0J( \omega_2)$, in which the spectral density function is given by~\cite{hwang,Halle,SI},
\begin{eqnarray}
    J(\omega) &=& \Gamma Re\Big[\frac{1+\frac{\sqrt{i\xi}}{4}}{1+\sqrt{i\xi}+\frac{4(\sqrt{i\xi})^2}{9}+\frac{(\sqrt{i\xi})^{3}}{9}}\Big]
\end{eqnarray}
where $\xi=\omega\tau_c$ with $\omega$ the relevant energy frequency (e.g. $\omega_0$
or $\omega_2$). $\alpha_0$ is a constant involving the nuclear properties of
the interacting system. $\Gamma$ is the coefficient determined by system parameters,
such as minimum distance between the electron and solvent spins, dipole-dipole coupling
strength and translational diffusion coefficient etc (see the SI~\cite{SI} for
details). In simple solvents of low viscosity such as protons in free water
molecules $D_T = D_w=2\times10^{-9}$ $m^2/s$~\cite{Krynicki} and for a minimal distance
of the electron and solvent spins of $b=0.5$ nm we find $\tau_c=b^2/D_T\sim0.1$ ns.
Then $\xi\ll1$ is always satisfied for weak and moderate magnetic fields (i.e. $B < 1$ T)
resulting in $\omega_S=(2\pi)16$ MHz for $B = 0.36$ T as shown by the blue dashed line
in Fig. \ref{J0}(a). Applying a MW driving fulfilling $\omega_E = \omega_S$ will
lead to no appreciable difference between flip-flip and flip-flop effects, as
$J\left(\omega_2\right)\cong J\left(\omega_{0}\right)\cong J(0)$ in general. This
indifference of the transition rates to the energy mismatch is the main reason
why resonance-based polarization schemes are not applicable to solutions with small
$\tau_c$. However, larger interaction correlation times achieved by increasing the
distance $b=5$ nm and decreasing the diffusion coefficient $D_T=10D_w$ to induce
$\tau_c=100$ ns, as shown in the purple dotted line in Fig. \ref{J0}(a)), demonstrate
a clear incline of net polariation towards resonance condition. Using the same
parameters discussed above, we now obtain $J(\omega_0)= J(0)\sim1$ and $J(\omega_2)\sim0$.
Therefore, it is possible to achieve a polarization transfer in solution with large
$\tau_c$ by matching a H-H resonant condition. It is interesting to note that the
relationship between our scheme and standard Overhauser cross-polarisation is similar
to the sideband resolved regime and the Doppler (unresolved) regime in laser
cooling of cold trapped ions (see SI~\cite{SI}). This results in a more efficient polarisation
transfer, which enables DNP via more distant electron spins, and with a weaker dipolar
coupling to the nuclear spins.


The steady state populations in the different nuclear spin states and the
resulting polarization can then be determined from a detailed balance analysis.
We find (we ignore here relaxation times, see SI~\cite{SI}
for a more detailed treatment)
\begin{eqnarray}
    P_s &=& -\frac{J(\omega_0)-J(\omega_2)}{J(\omega_0)+C_0J(\omega_S)+J(\omega_2)},
\end{eqnarray}
in which $C_0=2\cot\varphi$ with $\cot\varphi=\epsilon/\Omega$ depending on the Rabi
frequency of the MW radiation $\Omega$ and detuning of the MW field from the
electron energy scale $\epsilon$. Here we assume that the electron spin is continuously
pumped into a polarized state. Fig. \ref{J0}(b) shows the dependence of the achieved
steady-state polarization on the detuning from the resonance condition and correlation
time. As we are using the solid effect, we tune the MW field to achieve resonant interaction
as $\omega_S= \sqrt{\epsilon^2+\Omega^2}=\omega_E=(2\pi)16$ MHz (see SI~\cite{SI}),
where we have chosen $\epsilon=\Omega=(2\pi)8\sqrt{2}$ MHz. As shown in Fig. \ref{J0} (a),
for a large correlation time satisfying $\tau_c>10$ ns, different rates between the flip-flop
and flip-flip transitions induce a high steady-state polarization, quite similar to solid-state mechanisms. However, contrary to the solid-state phase, we can see that high steady-state
polarization is achieved for a wide range of detuning from the resonance, as shown in Fig.
\ref{J0}(b). For example, for $\tau_c=100$ ns, we observe a high steady-state polarization
for a detuning as large as $\Delta'\sim(2\pi)10$ MHz. This is a unique characteristic of
our theory in the liquid phase and different from the solid-phase in which polarization
transfer only occurs when the detuning is comparable or smaller than the effective collective
flip-flop coupling between the electron spin and nuclear spins~\cite{chen,London}. Thus,
this polarization by cross-relaxation requires an inclination towards resonance, but
the resonance condition does not have to be matched exactly.

\textit{Optical SPRINT with nanodiamonds--} To model the dynamics of nuclear spin polarization
transfer from NV centers in nanodiamonds to the protons in water, we consider a setup of
Fig. \ref{setup}(a). The solvent spins of the target molecules are pumped through a flow cell
containing a hyperpolarization region in which nanodiamonds are immobilized in a hydrogel
layer (see Fig. \ref{setup}), causing a dipolar magnetic interaction between the NV spins
and their surrounding solvent spins with a characteristic correlation time $\tau_c$. Here the
electron spins of the NV centers will be strongly polarized by optical pumping and the
hydrogel provides a method for increasing the correlation time by reducing the
diffusion rate of water molecules by a factor of $10^2-10^3$ via controlling the mesh
sizes and types of hydrogels~\cite{Shapiro}.

A DC magnetic field $B \cong 0.36$ T whose direction defines the z-axis of the
laboratory frame is applied to the system. We assume $\gamma_{e}B\gg D$ where $\gamma_{e}B$
denotes the NV centre Larmor frequency and $D$ its zero-field splitting. As discussed
in Ref.~\cite{chen}, in this regime the quantisation axis of all NV centres
is along the magnetic field direction, and the orientation of the symmetry axis of the NV center relative
to the external magnetic field is uniformly distributed over the unit sphere. Here $\theta$
denotes the angle between the NV center symmetry axis and the magnetic field axis. The
random orientations of the NV centers causes two difficulties for the polarization transfer
from the NV spin to the nuclear spins. First, high optical polarization of the NV center
spins is only achieved for the NV orientation near $\theta = 0^\circ$ and $\theta =90 ^\circ$
by using 532 nm green laser illumination and, secondly, there's a variation of the energy
splitting with $\theta$ (see SI~\cite{SI}).

The rate of the polarization transfer is given by $W=W_0-W_2 = \alpha_0(J(\omega_0)-J(\omega_2))$~\cite{SI}.
Assuming the magnetic field is given by
$B=0.36$ T, $\omega_S=(2\pi)16$ MHz, for utilizing the effect of resonance in the current
setup, we tune the MW frequency to achieve resonant interaction for $\theta = 90^{\circ}$,
and $\varepsilon_0=\varepsilon(90^\circ)=\Omega=(2\pi)8\sqrt{2}$ MHz for matching a H-H
resonance. Due to the strong dependence of the energy splitting between the spin levels
on $\theta$, for each NV spin with a specific $\epsilon(\theta)$, there is a
detuning from resonance which leads to $\Delta'$ in Fig. \ref{J0}(b).
The reason that we focus here on the near resonant case around $\theta =90 ^\circ$
can be found in the observation that the same detuning leads to more NV spins to be
involved in the $\theta=90^\circ$ case as compared to the $\theta=0^\circ$ case (e.g.
$\Delta'<(2\pi)10$ MHz involves 5\% NV spins in nanodiamonds in the former as compared
to just 0.1\% NV spins for the latter \cite{chen}.
We estimate the efficiency of our scheme by calculating the average polarization rate of
the solvent spins $\overline{ W}_{eff}= S^{-1}\int_{S}WP_C\cos\varphi dS$, in which
$P_C\cos\varphi$ is the initial polarization of the corresponding NV spin according to
the solid effect polarization mechanism~\cite{wenckebach2008solid} and $S$ is the solid
angle covered.

\begin{figure}
\center
\includegraphics[width=2.4in]{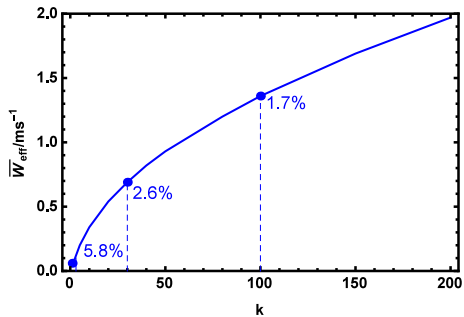}
\caption{The averaged polarization rate of the solvent spins $\overline{ W}_{eff}$. $b=5$ nm ND radius was assumed. $k=D_w/D_T$ with $D_w=2\times10^{-9}$ $m^2/s$ the diffusion coefficient in free water at room temperature and $D_T$ is the decreased translational diffusion coefficient. The percent of involved NV spins accounting for 90\% of the polarization transfer is given for three different points.
}
\label{Weff}
\end{figure}

Consider continuous optical pumping as well as optical initialization of the NV spins
for $\theta=90^\circ$, where $P_C\simeq0.5^2=0.25$. The dependence of the average polarization
rate of the solvent spins for nanodiamonds of $b=5$nm radius on the ratio of the decreased
translational diffusion $k$ is shown in Fig. \ref{Weff}. 
Slow translational diffusion of the solvent molecules affects the polarization transfer in
two competing ways: (i) increasing the correlation time to enhance the polarization transfer
and (ii) decreasing the number of the involved NV spins contributing to the polarization
transfer (as slower diffusion leads to smaller accessible detuning from resonance for our
scheme). Fig. \ref{Weff} predicts that the former will dominate
(as expected, since the first effect is linear and the second sub-linear). Thus, a fast
polarization transfer can be achieved via a decrease of the diffusion rate. For example,
when the size of the nanodiamond is $b=5$ nm, $\overline{ W}_{eff}\approx0.04$
ms$^{-1}$ for $k=1$ and $\overline{ W}_{eff}\approx1.4$ ms$^{-1}$ for $k=100$.



The proposed setup is feasible with current experimental technology. Over the
past decade, several experiments have realised polarization of flowing solvents by
immobilized electron spins such as TEMPO and radicals in hydrogel layers~\cite{mccarney2008spin1,mccarney2008spin2,ebert2012mobile} for $B\simeq0.35$ T. Furthermore, it has been demonstrated that one can use functional
groups on the diamond surface to replace these radicals with nanodiamonds in the
hydrogel~\cite{HuangH,Haartman}. The nanodiamonds are immobilized in the hydrogel
and it is not necessary to consider rotational diffusion. While the implementations
employing either radicals or nanodiamonds are similar in spirit, it
is crucial that the transparency of the hydrogel allows for the optical polarization
of nanodiamonds which has the potential to lead to a 300-fold increase of the achievable
nuclear polarisation over the implementations using radicals.

To estimate the total polarization of the solvent spins, we use an approximate formula
for the steady state bulk nuclear spin polarization of the solvent neglecting
polarization diffusion,
\begin{eqnarray}
    P_{lim} &\approx&\frac{N_e}{N_p}\overline{W}_{eff}T_p.
\end{eqnarray}
Here $\frac{N_e}{N_p}$ is the ratio of the number of the NV spins to proton spins in the polarization region
(see SI~\cite{SI}), and $T_p$ is the smaller one of the
times that the solvent contacting time with the hydrogel matrix or the solvent relaxation time. Suppose
the flow channel is composed of a tube of a diameter of 1 mm, the size of the hydrogel layer
is 1 mm and the resulting volume is filled with nanodiamonds of 10nm diameter such that they
account for 12\% of the total volume. Assuming $k=100$, we choose the flow
rate such that a molecule of water passes through the hydrogel layer in 1 second, so that
$T_p \sim T_1 \sim 1 s$, namely $v = 10^{-3}$ m/s (this low flow rate is easily produced via
commercial pumps, and negligibly affects the molecular dynamics within $\tau_c$). Given the
density of protons in free water $\frac{200}{3}$ nm$^{-3}$, we obtain $N_e/N_p\sim4\times10^{-6}$,
$\overline{ W}_{eff}\approx1.4$ ms$^{-1}$, $T_p\approx1$ s to arrive at a polarisation of
$P_{lim}\approx0.6\%$ for the solution. This polarization, achieved for a volume of $1\mu l$ in 1s
is roughly $4700$ times the thermal polarization and would already suffice for detection in a
commercial NMR scanner. It is interesting to note that a recently proposed setup for polarization
of fluids in a microfluidic diamond channel achieves similar degree of polarization, but due to
the limitations on the microfluidic channel size in the proposed protocol, it requires over
$1000$ parallel channels for polarizing 1$\mu$l of solvent within 1 second~\cite{Abrams}.

The volume of optimally polarized fluid that can be generated per unit time in a flow
channel largely depends on the thickness of the hydrogel layer, as faster flow velocities
for thicker layers are easily accessible (to maintain $T_p \sim T_1$). One of the limitation
in our setup is the laser intensity required to optically polarize the NV centre spins
within 10-100 $\mu$s. Experiments performed in bulk diamond have shown that typically
available lasers are sufficient for efficiently polarizing NV centers in 1 mm$^3$ volume, though optical absorption by the NV centers would limit the efficient optical polarization for $> 1$ mm thickness.
Furthermore, due to the small nanodiamond radius, Mie scattering does not limit the optical
penetration depth in the 1 mm volume for the nanodiamond concentration in the proposed setup~\cite{SI},
and consider a flow rate of 1mm/s, the temperature increases due to heating is expected to be limited
to only a few degrees. A more stringent limitation on the channel diameter comes from
the microwave attenuation. 1 mm diameter capillaries are typically used for solutions due to
the microwave absorption by water, which limits the microwave power for larger channels~\cite{nesmelov2004aqueous}.
Another important factor for the polarization transfer is the relaxation time $T_1$ of the
protons in the hydrogel layer, as discussed in Ref.~\cite{Shapiro}, this would depend on the
material and mesh size of the gel, but can be assumed to be 0.3-2s.

While we have demonstrated theoretically that our setup in combination with SPRINT
can achieve significant hyperpolarization of water, we would like to stress that a wide
variety of molecules can be polarized with the same methodology, including those with
nuclear spin species different from hydrogen. Of particular interest here are $^{13}C$
nuclear spins for example in $^{13}C$ pyruvic acid or $^{13}C$ glucose which have
applications in biomedical imaging and can be polarized via the same setup. The decrease in
polarization rate due to the fact that the gyromagnetic ratio of carbon is four times
smaller than that of protons is typically offset by the longer nuclear spin relaxation time.
Larger molecules could also potentially be polarized in our setup, especially as the interaction
correlation time with these molecules is naturally longer. In this scenario the properties
of the hydrogel would have to be tuned to avoid a dramatic reduction of the molecules'
diffusion coefficient.

SPRINT can potentially be very effective for non-optical DNP systems, not involving
nanodiamonds, for example for DNP in high magnetic fields and/or with biomacromolecules
(e.g. proteins). In these systems, if the interaction correlation time between the polarizing
electron (e.g. in a radical) and the molecule can be increased, it would be relatively
straightforward to apply our scheme. We anticipate that with minor modifications,
resonance-inclined transfer could be applied in combination with other solid-phase schemes
such as the cross effect.

\textit{Conclusion --} We developed SPRINT, a resonance-inclined mechanism for the polarization
of nuclear spins for large interaction correlation times in a solution. Importantly, this is a
complementary method to the Overhauser effect which is most efficient in the extreme narrowing regime.
Due to molecular motion, our scheme is tolerant to a relatively
large deviation from the resonant frequency. Furthermore, we propose a polarization setup using
NV centres in nanodiamonds held in a hydrogel inside a flow cell which combines the advantages
of this scheme with optical electron polarization. Under realistic experimental conditions, a
polarization enhancment of 4700 times is observed. Finally, we predict that our resonance-inclined
scheme could be used in a wide variety of DNP realizations, especially with biomacromolecules or
in large magnetic fields.

\textit{Acknowledgements --} We thank Yuzhou Wu, Pelayo Fernandez Acebal, Oded Rosolio
and Martin Bruderer for the insightful discussions. This work was supported by an Alexander
von Humboldt professorship, the ERC Synergy Grant BioQ, the ERC POC grant NDI, an IQST PhD Fellowship, the EU projects
SIQS, DIADEMS and HYPERDIAMOND as well as the DFG CRC TR 21.

\newcommand{\beginsupplement}{%
        \setcounter{table}{0}
        \renewcommand{\thetable}{S\arabic{table}}%
        \setcounter{equation}{0}
        \renewcommand{\theequation}{S\arabic{equation}}%
        \setcounter{figure}{0}
        \renewcommand{\thefigure}{S\arabic{figure}}%
     }
\beginsupplement
\widetext
\clearpage
\section*{Supplementary Information to the manuscript\\
``Resonance-inclined optical nuclear spin polarization of liquids in diamond structures"}
\subsection{Mechanism}
\subsubsection{Basic Hamiltonian}
For illustration of the basic idea, let us first consider a general system that is composed of an immobilized electron spin $1/2$ (notice that the NV spin $S=1$ could be simplified to be a two-level system which we will discuss later) and all its nearby diffusing solvent spins $I=1/2$. The Hamiltonian for such a system is then
 \begin{eqnarray}
    H &=& \omega_{0s}\sigma_{z}+2\Omega(\cos\omega_{dr}t)\sigma_{x} +\sum_{n=0}^{N_p}\gamma_{n}BI^{n}_{z}+ \sum_{n=0}^{N_p} g_n\Big[\vec{\sigma}\vec{I^n}-3(\vec{\sigma}\cdot\vec{e}_{nr})(\vec{I^n}\cdot\vec{e}_{nr})\Big],
\end{eqnarray}
in which $\omega_{0s}$ represents the Zeeman splitting of the electron spin and the last term is the magnetic dipole-dipole interaction between the electron and nuclear spin. The length dependant factor $g_n$ and orientation of the inter-spin vector $\vec{e}_{nr}$ is modulated by the translational diffusion coefficient $D_T$, the rotational diffusion coefficient of the electron spin is not considered due to its immobilization in our setup. $\omega_{dr}$ is the driving microwave field of frequency, $\Omega$ is the Rabi frequency and $\gamma_n$ are the nuclear gyromagnetic ratio. Assuming a point-dipole interaction, neglecting the contact term and moving to a frame rotating with the drive frequency, we obtain
\begin{eqnarray}
    H' &=& \epsilon\sigma_{z}+\Omega\sigma_{x}+\sum_{n=0}^{N_p}(\sigma_z\cdot\mathbf{A_n}\cdot\vec{I}^n
    + \gamma_{n}BI^n_{z}),
\end{eqnarray}
where $\epsilon=\omega_{0s}-\omega_{dr}$, $\mathbf{A_n}=g_n\sqrt{1+3(e^z_{nr})^2}\vec{h}_n=A_n\vec{h}_n$ is the hyperfine interaction tensor for the nuclear spin, with $g_n=\frac{\mu_0}{4\pi}\frac{\gamma_e\gamma_n}{r_n^3}$ with $r_n=|\vec{r_n}|$ denoting the distance from electron spin to a nuclear spin. $\vec{h}_n$ determined by $\vec{e}_{nr}$: $h^x_n=3e^{x}_{nr}e^{z}_{nr}/\sqrt{1+3(e^{z}_{nr})^2}$, $h^y_n=3e^{y}_{nr}e^z_{nr}/\sqrt{1+3(e^z_{nr})^2}$, and $h^z_n=(3(e^z_{nr})^2-1)/\sqrt{1+3(e^z_{nr})^2}$. We also define $e^z_{nr}=\cos\theta'_n$, $e^x_r=\sin\theta'_n\cos\varphi'_n$ and $e^z_{nr}=\sin\theta'_n\sin\varphi'_n$. By redefining the parameters we get:
\begin{eqnarray}
    H_B = \omega_{E} \sigma_{\tilde z}+\sum_{n=0}^{N_p}[ \omega_SI^n_{z'}+(\sigma_{\tilde x}\sin\varphi + \sigma_{\tilde z}\cos\varphi)\cdot(a_{x'_n}I^n_{x'}+a_{z'_n}I^n_{z'})],
 \end{eqnarray}
in which $\omega_S=\gamma_{n}B$, the angle $\varphi$ and $\omega_{E}$ are determined by
\begin{displaymath}
    \sin\varphi=\frac{\Omega}{\sqrt{\epsilon^2+\Omega^2}}, \ \ \ \omega_{E}=\sqrt{\epsilon^2+\Omega^2}, \ \ \ \omega_{E} \sigma_{\tilde z}=\Omega\sigma_{ z}+ \epsilon\sigma_{ x} \nonumber
\end{displaymath}
with $a_{x'_n}=3g\cos\theta'_n\sin\theta'_n e^{i\phi'_n}$ and $a_{z'}=g(3\cos^2\theta'_n-1)/r_n^3$ in which $g=\frac{\mu_0\hbar\gamma_e\gamma_n}{4\pi}$. Note that we consider the regime where the electron frequency $\omega_{0s}$ is large enough to fulfill $\omega_{0s}\tau_c > 1$, i.e. outside the so-called ``extreme-narrowing" limit. This condition allows us to disregard fast-oscillating terms when applying the MW driving.

We can rewrite the Hamiltonian as
\begin{eqnarray}
    H'_B &=& \omega_{E} \sigma_{\tilde z}+\sum_{n=0}^{N_p}\omega_{S}I^n_{z'} +(\sigma_{\tilde x}\sin\varphi + \sigma_{\tilde z}\cos\varphi)\cdot[3g\sqrt{\frac{8\pi}{15}}Y^1_2(\theta'_n,\phi'_n)I^n_{x'}+g\sqrt{\frac{16\pi}{5}}Y^0_2(\theta'_n,\phi'_n)I^n_{z'}],
    \label{HB}
 \end{eqnarray}
with the spherical harmonics $Y^1_2(\theta'_n,\phi'_n)=-\frac{1}{2}\sqrt{\frac{15}{2\pi}}\cos\theta'_n\sin\theta'_n e^{i\phi'_n}$ and $Y^0_2(\theta'_n,\phi'_n)=\frac{1}{4}\sqrt{\frac{5}{\pi}}(3\cos^2\theta'_n-1)$. The Hartmann-Hann resonant condition is given by $\omega_E=\omega_S$.

\begin{figure}
\center
\includegraphics[width=2.8in]{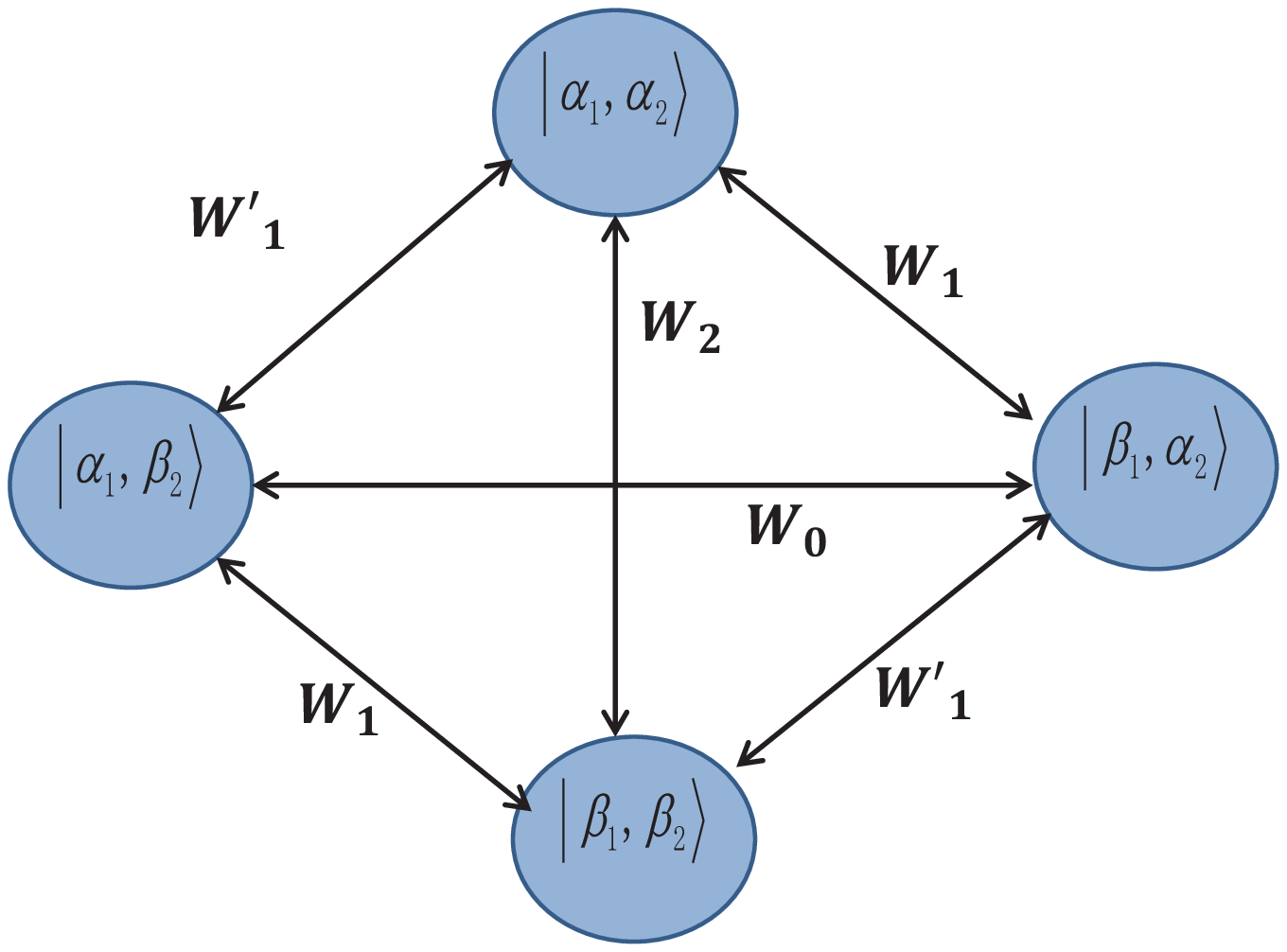}
\caption{Transition probabilities between the eigenstates of the spin operators.}
\label{Tp}
\end{figure}
\subsubsection{Transition probability}
In first order in perturbation theory the transition probability is given by $W_{ij}=\frac{1}{t} \bigg \langle |\int_0^t\langle i|H'_B(t')|j\rangle e^{i\omega_{ij}t'}dt'|^2 \bigg \rangle_{noise}$ we readily arrive~\cite{SolomonS}
\begin{eqnarray}
W_{ij}&=&\frac{1}{t} \bigg \langle \int_0^t  \int_0^t \langle i|H'_B(t')|j\rangle  \langle j \vert H'_B(t')\vert i \rangle e^{i\omega_{ij}(t'-t'')}dt' dt''\bigg \rangle_{noise} \nonumber\\
&=&\frac{1}{t} \int_0^t  \int_0^t  \bigg \langle \langle i|H'_B(t')|j\rangle  \langle j \vert H'_B(t')\vert i \rangle \bigg \rangle_{noise} e^{i\omega_{ij}(t'-t'')}dt' dt'', \nonumber\\
\end{eqnarray}
In principle this result is valid for short times as long as $e^{- W_i t} \ll 1,$ for longer times perturbation theory is invalid. However, this result is still valid for longer times in the limit $g \tau_c/r^3 \ll 1$ and $t \gg \tau_c$, which is correct in the regime of this work. In this regime and in the assumption of Gaussian noise we get two integrals which are connected by the fluctuation dissipation theorem for which the imaginary part is  just the power spectrum of the noise, or alternatively,
\begin{eqnarray}
    W_{i}(\omega_{i}) = \alpha_iRe[\int_{-\infty}^{+\infty}G_i(t')e^{i\omega_{i}t'}dt']=\alpha_i J(\omega_{i}),
    \label{Wi}
 \end{eqnarray}
 in which $\alpha_0=\alpha_2=\frac{3\pi g^2\sin^2\varphi}{10}$, $\alpha_1=\frac{3\pi g^2\cos^2\varphi}{5}$, $\omega_{0}=\omega_{E}-\omega_S$, $\omega_{2}=\omega_{E}+\omega_S$ and  $\omega_{1}=\omega_{S}$. All the related transitions are presented in Fig. \ref{Tp}, in which $|\alpha_1\rangle$, $|\beta_1\rangle$, $|\alpha_2\rangle$ and $|\beta_2\rangle$ are the eigenstates of electron and nuclear spins. $G_i(t')$ is the space-dependent time-correlation function for dipolar interaction between spins NV and proton, which is expressed by
 \begin{eqnarray}
   G_{i}(t') &=& \int\int N_pP(\vec{r_0})P(\vec{r_0}|\vec{r_1},t')\frac{Y^i_2(\theta^0,\phi^0)Y^i_2(\theta^1,\phi^1)}{r_0^3r_1^3}d\vec{r_0}d\vec{r_1},\nonumber
 \end{eqnarray}
in which $i=0,1$ and $G_{2}(t')=G_{0}(t')$. Here $P(\vec{r_0})=1/V$ with the volume $V$. $P(\vec{r_0}|\vec{r_1},t')$ is the well-known solution of the diffusion equation. As shown in Eq. (\ref{Wi}), the spectral density function is the real part of the Fourier-Laplace transform of the corresponding time autocorrelation function. Consider the diffusion coefficient $D_T$ in the whole space, one has \cite{Freed,HalleS}

 \begin{eqnarray}
   J(\omega) &=& \Gamma Re\Bigg[ \frac{1+\frac{\sqrt{i\xi}}{4}}{1+\sqrt{i\xi}+\frac{4(\sqrt{i\xi})^2}{9}+\frac{(\sqrt{i\xi})^{3}}{9}}\Bigg],
   \label{ps1}
 \end{eqnarray}
in which the coefficient $\Gamma=\frac{8n_p}{27D_Tb}$. Here $n_p$ is the proton number density, $D_T$ is the translational diffusion coefficient in hydrogel, $b$ is the closest distance between NV center and protons, and
\begin{displaymath}\xi=\frac{\omega b^2}{D_T}=\omega\tau_c.\end{displaymath}
It is interesting to know, the principle part of this integral will induce a two body shift: $(\sigma_z+1)(\sigma_z-1) + (\sigma_z-1)(\sigma_z+1),$ which is proportional to the power spectrum of $\omega_{diff}$ and $(\sigma_z+1)(\sigma_z+1) + (\sigma_z-1)(\sigma_z-1),$ which is proportional to the power spectrum of $\omega_{sum}.$ This term will create an energy shift if the NV which depends on the polarisation which for high polarizations should be compensated  by changing the drive.

A detailed balance between the populations in the different energy levels yields the steady-state polarization of solvent nuclear spins as ~\cite{SolomonS}
 \begin{eqnarray}
   P_s &=& -\frac{W_0-W_2}{W_0+2W_1+W_2+\frac{1}{T_1}}=-\frac{\alpha_0J(\omega_{0})-\alpha_2J(\omega_2)}{\alpha_0J(\omega_0)+2\alpha_1J(\omega_{1})+\alpha_2J(\omega_2)+\frac{1}{T_1}},
 \end{eqnarray}
in which $T_1$ is relaxation time of the NV spins. By ignoring $T_1$, we can achieve Equation (2) in the main text and $C_0=2\alpha_1/\alpha_0$. The rate of the polarization transfer is found to be given by
 \begin{eqnarray}
   W &=&W_0-W_2=\alpha_0J(\omega_0)-\alpha_2J(\omega_2).
 \end{eqnarray}
Notice that $W_0$ and $W_2$ are the transition rates of the flip-flop and flip-flip transitions, which are defined in the main text, respectively. The polarization built up depends on the imbalance between spectra density functions $J(\omega_{0})$ and $J(\omega_{2})$. 

The rate equation for the polarization transfer is in the form \cite{AbramsS}
 \begin{eqnarray}
  \frac{d\langle I\rangle}{ d t}=-\frac{\langle I\rangle}{T_1}-\frac{N_e}{N_p}W\langle\sigma \rangle,
 \end{eqnarray}
in which $\langle\sigma \rangle$ corresponding to the effective initial polarization of the electron spins. Therefore, when $\frac{d\langle I\rangle}{dt}=0$, in the stationary state, the formula for bulk nuclear spin polarization takes the approximate form,
\begin{eqnarray}
   P_{lim} &=&\frac{N_e}{N_p}WT_p\langle\sigma \rangle,
 \end{eqnarray}
in which $\frac{N_e}{N_p}$ is the ratio of the number of the NV spin and protons, and for solid effect the effective polarization which could be transferred to the nuclear spins is $P_C\cos\varphi$ with $P_C$ the optical polarization of the electron spin~\cite{wenckebach2008solidS} by continual optical pumping.


\begin{figure}
\center
\includegraphics[width=2.6in]{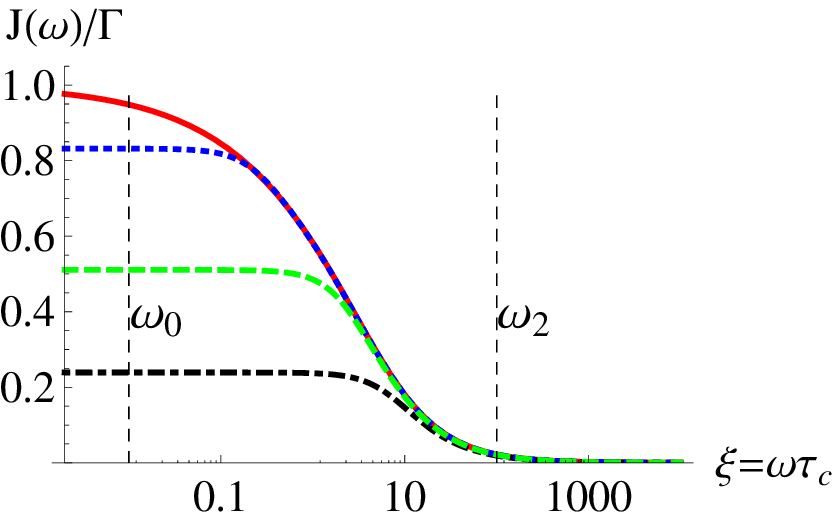}
\caption{Normalized spectra density function $J(\omega)/\Gamma$ with $\tau_c=10$ ns as a function of $\xi$ for selected absorbing boundary, from top to bottom referring to $b=5$ nm, $c=\infty$ (red solid), 50 (blue dotted), 17 (green dashed) and 10 (black dot-dashed) nm. Here we define $\omega_{E}-\omega_S=\omega_0$ and $\omega_{E}+\omega_S=\omega_2$.}
\label{J0A}
\end{figure}

\subsubsection{Connection to cooling}
Polarization is strongly connected to cooling as in both cases the aim is to prepare the system in a defined state.
Thus, as the field of laser cooling is very advanced it is interesting to compare the presented theory with the theory of laser cooling. Cooling is of atoms, ions~\cite{Chu,Cohen,Phillips} and mechanical oscillators~\cite{Igancio,Florian} is divided into two main regimes. The unresolved limit, or Doppler cooling~\cite{Hansch} and the resolved regime or sideband cooling~\cite{Wineland1,Wineland2,Stenholm} and more advanced methods as dark state cooling~\cite{Aspect,Dum,Giovanna,Us}.  In this work we point out for the first time the availability of the regime in which the sideband are resolved for dynamical nuclear polarization.

In both cases the final temperature and the rate are derived from the power spectrum of the noise~\cite{Cirac_Cooling}.
For laser cooling the noise originates from the interaction of the atoms with the vacuum while in our case there are two sources of noise.
The first one is the noise due to the random feature of the interaction between the NV and the nuclear spins. The second one is the vacuum which induces polarization via the lasers exactly as in the atomic version. In this work we assume that the second one is less significant which means that NV decay time ($T_1$) is longer than the decay time induced by the interaction noise.

In our case we have derived directly the rates from the first order term in perturbation theory while for laser cooling the noise and the auxiliary system are being adiabatically eliminated in order to derive a new Master equation for the system alone. The same could be done here. The adiabatic elimination could be done in two stages. In the first stage the noise should be eliminated to derive a Master equation of the nuclear spins and the NV. The adiabatic elimination could be done in the weak coupling limit and assuming Gaussian noise~\cite{avron,Spohn,alicki}. At the second stage the NV should be eliminated. In case of no nuclear driving only the the Lindbladian terms and an energy shift survive.

Thus, the final results are readily derived from the power spectrum of the noise (Eq. (\ref{ps1})), which could also be written as:
\begin{equation}
J(\xi)= \Gamma\frac{81 \left(2 \xi +5 \sqrt{2} \sqrt{\xi }+8\right)}{8 \left(4 \sqrt{2} \xi ^{5/2}+27 \sqrt{2} \xi
   ^{3/2}+\xi ^3+16 \xi ^2+81 \xi +81 \sqrt{2} \sqrt{\xi }+81\right)}.
   \end{equation}
The two interesting limits are derived in $\xi \rightarrow 0,$ which corresponds to the Doppler limit and the limit $\xi \gg 1$ which corresponds to the resolved sideband limit. In the resolved  regime, the power spectrum is $\frac{J(\xi)}{\Gamma}=\frac{81}{4 \xi ^2}+O\left(\left(\frac{1}{\xi }\right)^{5/2}\right)$ and thus the final occupation is $1-\frac{81}{2 \xi ^2}+O\left(\left(\frac{1}{\xi }\right)^{5/2}\right),$ which deviates from full polarization with the scaling of  $ \bigg(\frac{1}{\omega_0 \tau_c}\bigg)^2.$ In comparison to atom or ion cooling in which the occupation behaves as $ \bigg(\frac{\gamma}{\nu} \Bigg)^2,$ where $\nu$  is the trap frequency which is analogous to $\omega_0$ and $\gamma$ is the decay rate which is analogous to the correlation time. The analogy between the correlation time and $\gamma$ is due to the fact that $\gamma$ sets the correlation time in the atomic case. It can be seen that the same scaling is achieved up to the proportionality coefficient.

 In the unresolved limit, i.e., the Doppler limit, the power spectrum behaves as $1-\frac{3 \sqrt{\xi }}{8 \sqrt{2}}+O\left(\xi ^{3/2}\right)$ and thus
 the final occupation is   $\frac{3}{4 \sqrt{2}} \sqrt{\omega \tau_c}.$ Due to the $\sqrt{\xi}$ scaling the effective temperature  is not only a function of the decay rate as is in the laser cooling case, but a function of the energy gap as well, meaning, the final temperature is
 $\sqrt{\frac{\omega}{\tau_c}}.$

\subsubsection{Restricted uniform diffusion with absorbing boundary}
Here we isolate the contribution of the polarization transfer from the solvent spins located in the shell $b<R<c$, in which $c$ is the variable radial distance from the center of the nanodiamond. Therefore, assuming absorption boundary condition at the radial distance $c$ from the center of the nanodiamond, we can have
 \begin{eqnarray}
   J_l(\xi) = Re\Big\{\frac{2 n_s}{\xi^2D_Tb}\Big[\frac{1-l^3}{3}-\frac{3(1+l^5)}{\xi^2}
   +\frac{1}{C_2}(\frac{6l^3}{\xi^4}-\frac{9C_1}{\xi^3}-\frac{l^4C_3}{\xi})\Big]\Big\},
 \end{eqnarray}
in which
 \begin{eqnarray}
   C_1 &=& i_2(\xi)k_2(\xi/l)-k_2(\xi)i_2(\xi/l),\\ \nonumber
    C_2 &=& i'_2(\xi)k_2(\xi/l)-k'_2(\xi)i_2(\xi/l),\\ \nonumber
     C_3 &=& i'_2(\xi)k'_2(\xi/l)-k'_2(\xi)i'_2(\xi/l),
 \end{eqnarray}
with $l=b/c$. Here $i_2(\xi)$ and $k_2(\xi)$ are the related modified spherical Bessel functions~\cite{HalleS}. To elucidate the resulting interplay of the molecular frequency and the spectrometer frequency, suppose $b=5$ nm, we also show the complete normalized frequency spectrum for selected absorbing boundary $c=\infty$, 50, 17 and 10 nm, see Fig. \ref{J0A}. It's also very important to notice that, in the low frequency limit, the normalized spectra density has substantial contribution from solvent spins far beyond the first hydration layer, which enhances the polarization transfer rate.

\subsection{Optical nuclear spin polarization with nanodiamonds}
\subsubsection{Random orientations of the NV spins in nanodiamonds}
\begin{figure}
\center
\includegraphics[width=2.1in]{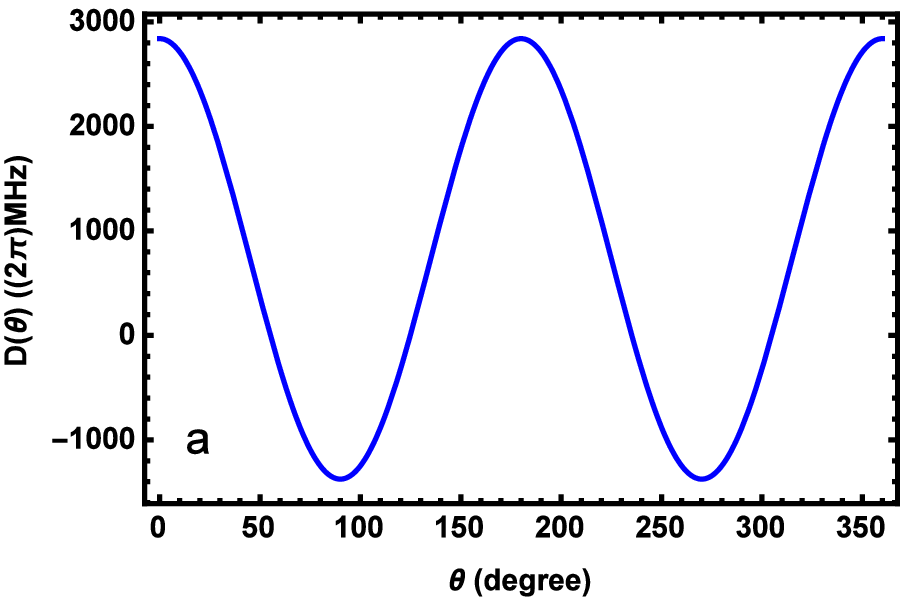}
\includegraphics[width=2.1in]{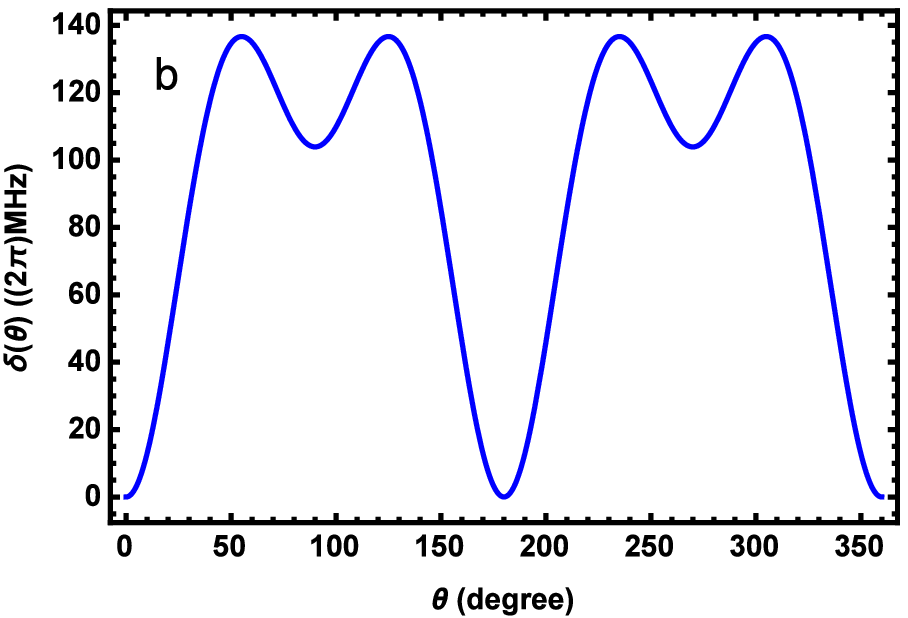}
\includegraphics[width=2.1in]{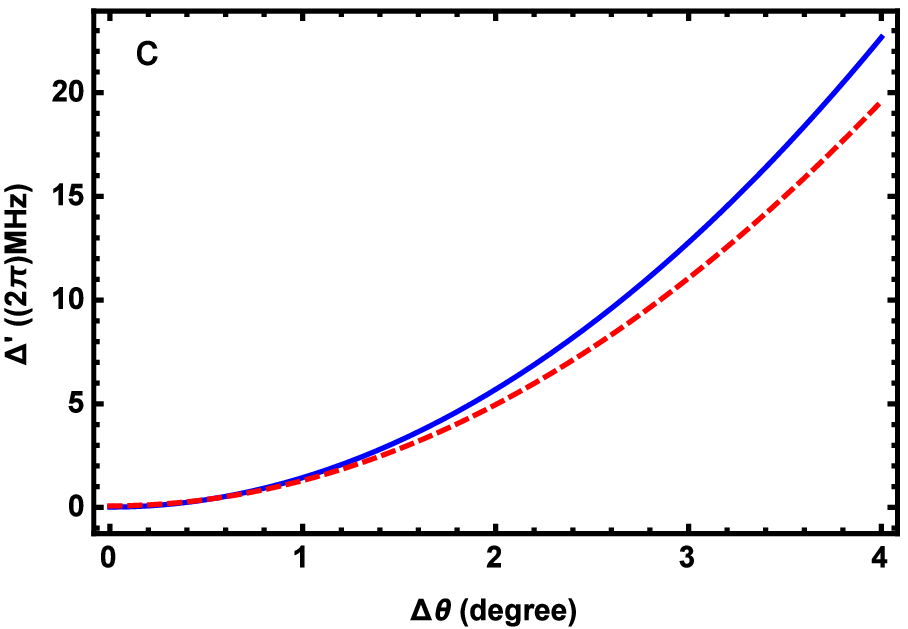}
\caption{a, Zero-field distribution $D(\theta)$ of the NV spins in nanodiamonds with $D=(2\pi)2.87$ GHz and $E=(2\pi)20$ MHz. b, The second order corrections which induce a energy distribution $\delta(\theta)$.  c, The detuning distributions $\Delta'=\epsilon-\epsilon_0$ vs the angle deviations for $\varepsilon_0=\epsilon(90^{\circ})$ presented by the red dashed line and $\epsilon_0=\epsilon(0^{\circ})$ by the blue solid line, respectively.}
\label{distribution}
\end{figure}

\textit{Zero-field and external magnetic field distribution}: In the laboratory frame the applied magnetic field defines the z-axis, the NV is placed at the origin of the coordinate system. The Hamiltonian of the electron spin is then~\cite{Chen2}
 \begin{eqnarray}
    H_{NV} &=& (\gamma_{e}B+\delta(\theta))S_{z}+D(\theta)S_{z}^{2}
\end{eqnarray}
in which
\begin{eqnarray}
D(\theta)=\frac{D(1+3\cos(2\theta))+3E(1-\cos(2\theta))}{4},\ \  \ \ \delta(\theta)=\frac{\gamma_e B|G_1|^2}{(\gamma_e B)^2-[D(\theta)]^2}+\frac{|G_2|^2}{2\gamma_e B},
\end{eqnarray}
with $G_1=\frac{(D-E)\sin\theta\cos\theta}{\sqrt{2}}$, $G_2=\frac{D+3E+(E-D)\cos2\theta}{4}$, zero-field splitting $D=(2\pi)2.87$ GHz and the local strain $E=(2\pi)20$ MHz. Clearly, the random orientations of the NV centers cause a variation of the zero-field splitting $D(\theta)$ across the entire interval $[-(2\pi)1.43\mbox{GHz},(2\pi)2.87 \mbox{GHz}]$ and $\delta(\theta)$ across the interval $[0 \mbox{MHz},(2\pi)140 \mbox{MHz}]$ as shown in Fig. \ref{distribution}a and b, respectively.

Considering the ground state, the states $m_s=0$ and $m_s=-1$ manifold form a two-level system, a microwave (MW) field of frequency $\omega_M$ is applied as a drive of the spin transitions $|-1\rangle \leftrightarrow |0\rangle$, $H_{dri}=\Omega_M S_x\cos\omega_M t$ ($\Omega_M$ is the Rabi frequency of the driving field and $\omega_M$ is its frequency). Working in a frame that rotates with the microwave frequency $\omega_M$, we find
\begin{eqnarray}
    H &=& \frac{\Omega}{2}(|0\rangle\langle-1|+|-1\rangle\langle0|) +\epsilon(\theta)|-1\rangle\langle-1|+(|-1\rangle\langle-1|)\cdot\mathbf{A}\cdot\vec{I}
    +\gamma_{n}BI_{z},
\end{eqnarray}
The detuning is given by $\epsilon(\theta)= D(\theta)-\gamma_{e}B-\delta(\theta)+\omega_M$, and $\Omega=\Omega_M/\sqrt{2}$. Representing the system with the electronic dressed states $|\pm\rangle=\frac{1}{\sqrt{2}}(|-1\rangle\pm|0\rangle)$ and defining the Pauli operator as $\sigma_{z}=\frac{1}{2}(|+\rangle\langle+|-|-\rangle\langle-|)$, $\sigma_{x}=\frac{1}{2}(|+\rangle\langle-|+|-\rangle\langle+|)$, then the polarisation transfer dynamics is described by
\begin{eqnarray}
\label{Htrans}
    H_{trans}&=& \Omega\sigma_{ z}+ \epsilon\sigma_{ x}+\omega_{S}I_{z'} + a_{z'} \sigma_{ z} I_{z'} + a_{x'}\sigma_{x} I_{x'}.
\end{eqnarray}
Here $\omega_{S}=\gamma_n\vec{B}-\vec{A}$, And we can rewrite this Hamiltonian as Equation (\ref{HB}). 
%

\textit{Optical initialization}:
A randomly oriented nanodiamond ensemble concerns the optical polarisation of electron spins of the NV center. For bulk
diamonds, the magnetic field can be aligned with the principal axis of the NV center and the
electronic spin of the NV center can be optically polarised to the state $|m_s=0\rangle$
by illumination with a $532$ nm green laser. However, although in the laboratory frame of reference a strong magnetic field is applied
along the $z$-direction, $\gamma_{e}B_{z}\gg D$, for an ensemble of randomly oriented
nanodiamonds, the NV centers will be optically
pumped to the state $|m_s=0\rangle_{\theta}$ that is defined by the relative orientation
of the NV center with respect to the externally applied magnetic field which defines the
laboratory frame~\cite{Chen2}. Here the misalignment angle between the NV-axis and
the magnetic field is $\theta$.

As discussed in Ref.~\cite{Chen2}, we have two coordinate systems which can be transformed into each other
and, employing of $ S_{z_{\theta}} = \cos\theta S_{z}-\sin\theta (\cos\phi S_{x}-\sin\phi S_{y})$,
we can express the eigenstate $|m_s=0\rangle_{\theta}$, i.e. in the lab frame,
\begin{equation}
\label{optical}
|m_s=0\rangle_{\theta} = \cos\theta|0\rangle + \frac{\sin\theta}{\sqrt{2}}
(e^{i\phi}|+1\rangle-e^{-i\phi}|-1\rangle).
\end{equation}
If $\theta$ is large, the eigenstate $|0\rangle$ of the NV center in the laboratory frame differs significantly from the zero-field eigenstate
$|0\rangle_{\theta}$ of the NV center. Hence optical initialisation of randomly oriented NV
centers lead to very different states depending on the orientation of the NV center.

Notice that for small misalignment between the NV center and external magnetic field,
the initialisation of the NV center is well approximated to $|0\rangle$. Indeed, for $\theta<10^\circ$, we can arrive significant polarisation of the electron spins along the quantisation axis defined by the external magnetic field $P_{NV}\simeq1$. For $\theta\in[80^\circ,100^\circ]$,
the population in state $|0\rangle$ is very small and the initial state is well approximated by
$\frac{1}{\sqrt{2}}(e^{i\phi}|+1\rangle-e^{-i\phi}|-1\rangle)$. In the subspace spanned by the states $\{|0\rangle, |-1\rangle\}$, the NV spin
is well-polarized in state $|-1\rangle$ with polarization $P_{NV}\simeq 0.5$. So consider a continual optical pumping, it's reasonable to have the initial polarization of the NV spin $P_C\simeq0.25$.

\textit{Detuning from the resonance}: MW field is applied to have $\epsilon_0=\Omega=(2\pi)8\sqrt{2}$ MHz which matches the resonant condition $\omega_S=\omega_E=(2\pi)16$ MHz. Since high polarization is achieved for $\theta<10^\circ$ and $\theta\in[80^\circ,100^\circ]$, assume two resonant points with $\epsilon_0=\epsilon(0^\circ)$ or $\epsilon_0=\epsilon(90^\circ)$, we calculate the detuning from the resonance which is defined as $\Delta'=\epsilon(\theta)-\epsilon_0$ and varies with the angle deviation from $0^\circ$ and $90^\circ$, respectively. As shown in Fig. \ref{distribution}c, if the angle deviation $\Delta\theta\sim4^\circ$, this detuning reaches $(2\pi)20$ MHz both for $\theta<10^\circ$ and $\theta\in[80^\circ,100^\circ]$.

\begin{figure}
\center
\includegraphics[width=2.6in]{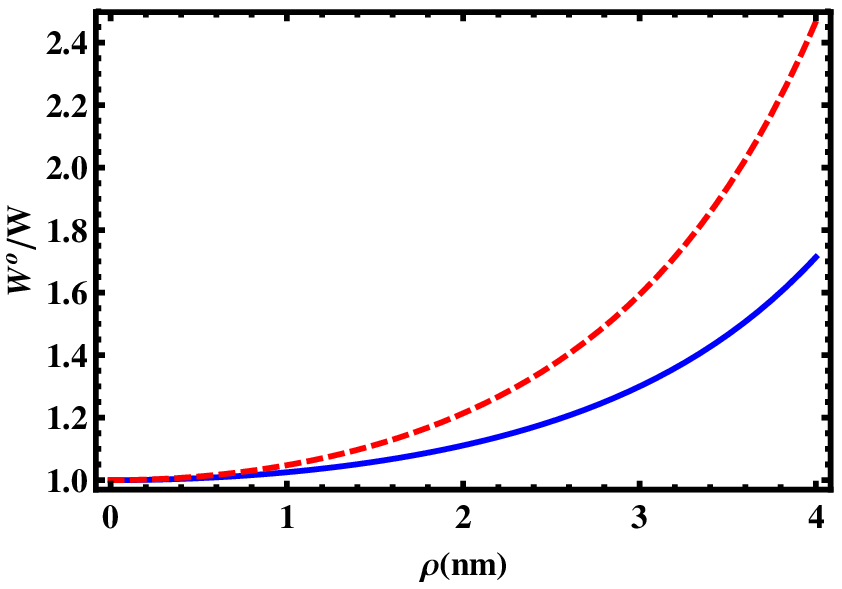}
\caption{The ratio of polarization transfer with and without eccentricity vs the distance from the center of the nanodiamond. The red dashed and blue solid line show the dependence of comparison of the polarization built up for  $\omega_S=(2\pi)16$ MHz, $\Delta_0=\Omega=(2\pi)8\sqrt{2}$ MHz, and $\Delta'=0$ and $\Delta'=(2\pi)10$ MHz, respectively.}
\label{off}
\end{figure}
\subsubsection{Off-center effect}
It's quite normal when the NV spin is not located in the center of the nanodiamond, so we intend to discuss the polarization transfer under the off-center effect. Suppose a NV center is located off-center and its distance from the center of the nanodiamond is $\rho$. Consider a series expansion in terms of the eccentricity parameters, we can include the off-center effect in the spectra density function which is given by \cite{Albrand}
 \begin{eqnarray}
 J^O(\xi)=[J(\xi)]\times[(1+\frac{1}{6}\sum_{L=1}^{\infty}(L+1)(L+2)(2L+3)\rho^{2L}\frac{Re[J^{L+2}(\xi)]}{J(\xi)}],
 \end{eqnarray}
where
 \begin{eqnarray}
J^{L+2}(\xi)=Re[\frac{2 n_s}{i\xi D_Tb}\{\frac{1}{2L-1}-\frac{L+1}{i\xi}[1+\frac{(L+1)K_{L+1/2}(\sqrt{i\xi})}{\sqrt{i\xi}K_{L-1/2}(\sqrt{i\xi})}]^{-1}\}],
 \end{eqnarray}
with a modified Bessel function $K_\mu(\sqrt{i\xi})$. Thus the rate of polarization transfer including off-center effect is given by $W^O=W^O_0-W^O_2=\alpha_0 (J^O(\omega_0)-J^O(\omega_2))$. Fig. \ref{off} illustrates the dependence of the polarization built-up on the eccentricity of the NV center. One can find that polarization built up only is enhanced by factor 1.7 ($\Delta'=0$ MHz) or 2.4 ($\Delta'=(2\pi)10$ MHz). It is manifested that the polarization transfer is benefited by the eccentricity due to the reduced distance between the NV and the neighboring nuclear spins, but this enhancement is not significant. The possible main reason is the long range nature of the spectral density function, which includes contributions from a large amount of the distant solvent spins.

\subsubsection{Optical penetration depth}
\begin{figure}
 	\center
 	\includegraphics[width=2.6in]{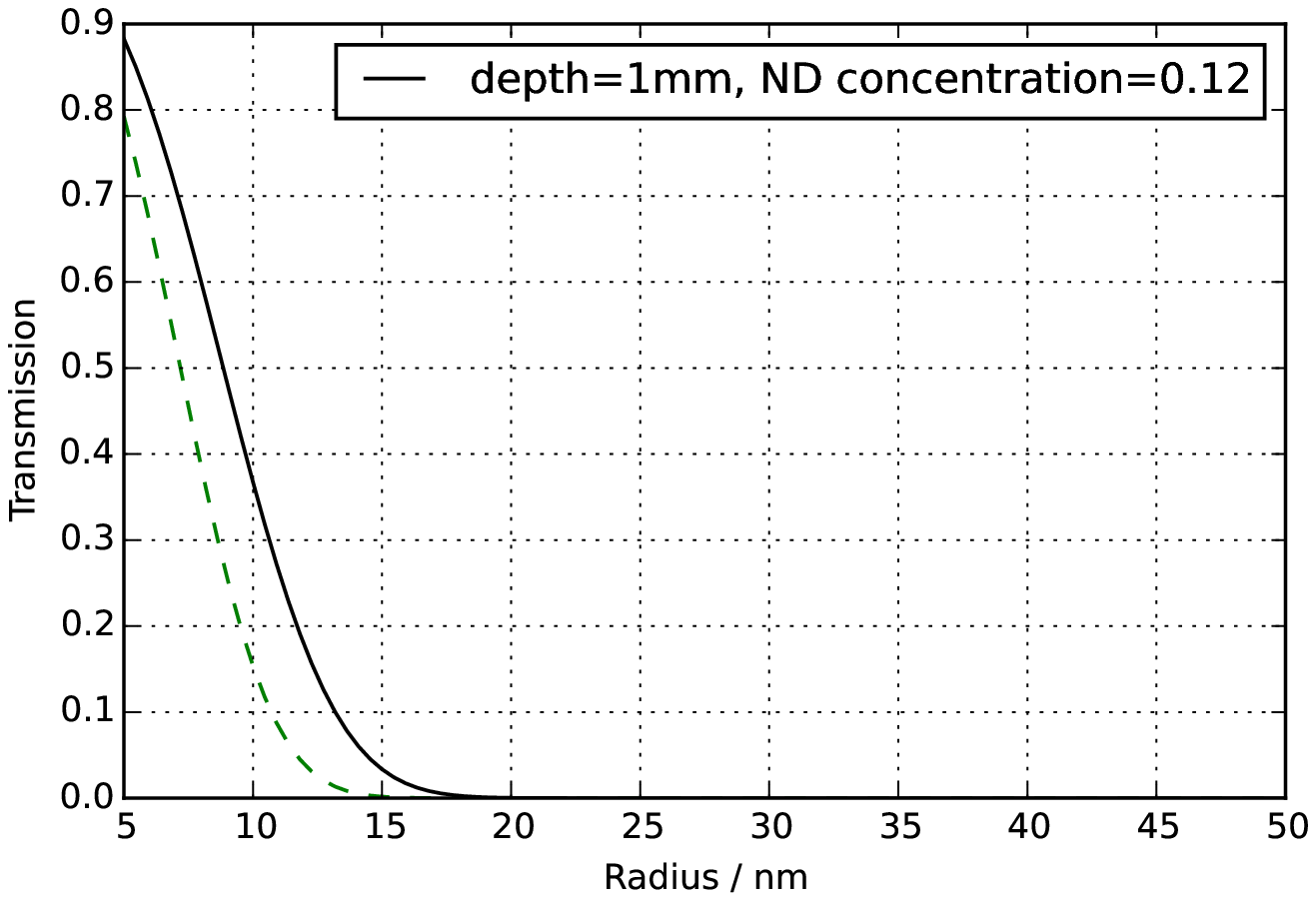}
 	\caption{The optical transmission through our setup dependence on the nanodiamond radius, taking into account Mie scattering from the nanodiamonds. We assume 1 mm thickness and 12\% ND concentration. The black solid and green dashed lines correspond to a refractive index of the medium in between the nanodiamonds of water and air respectively.}
 	\label{opt}
\end{figure}

There are two main limitations on the penetration depth when applying optical irradiation to the polarisation structure of the hydrogel containing imobilized diamond: scattering due to nanodiamonds and optical absorption (mainly by NV centre spins). Taking into account the optical cross section of the NV centres and the relatively low NV concentration in this setup we expect the optical penetration depth by the NVs to be a little larger than 1 mm. The optical scattering due to nanodiamonds could become an issue which limits the total penetration depth. As the nanodiamond size is small compared to the optical wavelength, but not negligibly so, we use Mie scattering to calculate the optical transmission through our setup (taking into account only the scattering from the nanodiamonds). It can be seen that the optical penetration decreases sharply with the nanodiamond radius, see Fig. \ref{opt}. However, for the ND size of interest $b=5$ nm, the scattering from NDs in 1 mm is almost negligible.

\end{document}